\documentclass[acmlarge]{acmart}
\AtBeginDocument{%
  \providecommand\BibTeX{{%
    \normalfont B\kern-0.5em{\scshape i\kern-0.25em b}\kern-0.8em\TeX}}}

\usepackage{listings}
\usepackage{multirow}
\usepackage{pifont}

\begin{document}

\title{A Checklist to Publish Collections as Data in GLAM Institutions}

\author{Gustavo Candela}
\email{gcandela@ua.es}
\orcid{0000-0001-6122-0777}
\affiliation{%
  \institution{University of Alicante}
  \city{Alicante}
  \country{Spain}
}

\author{Nele Gabriëls}
\email{nele.gabriels@kuleuven.be}
\orcid{0000-0002-0325-3228}
\affiliation{%
  \institution{KU Leuven Libraries}
  \city{Leuven}
  \country{Belgium}
}

\author{Sally Chambers}
\email{Sally.Chambers@kbr.be}
\orcid{0000-0002-2430-475X}
\author{Thuy-An Pham}
\email{thuy-an.pham@kbr.be}
\affiliation{%
  \institution{KBR, Royal Library of Belgium}
  \city{Brussels}
  \country{Belgium}
}

\author{Sarah Ames}
\email{S.Ames@nls.uk}
\affiliation{%
  \institution{National Library of Scotland}
  \city{Edinburgh}
  \country{Scotland}
}

\author{Neil Fitzgerald}
\orcid{0000-0002-0602-9993}
\email{Neil.Fitzgerald@bl.uk}
\affiliation{%
  \institution{British Library}
  \city{London}
  \country{United Kingdom}
}

\author{Katrine Hofmann}
\email{khg@kb.dk}
\author{Victor Harbo}
\email{vhol@kb.dk}
\affiliation{%
  \institution{Royal Danish Library}
  \city{Aarhus}
  \country{Denmark}
}

\author{Abigail Potter}
\email{abpo@loc.gov}
\author{Meghan Ferriter}
\email{mefe@loc.gov}
\author{Eileen Manchester}
\email{ejakeway@loc.gov}
\affiliation{%
  \institution{Library of Congress}
  \city{Washington}
  \country{United States of America}
}

\author{Alba Irollo}
\email{alba.irollo@europeana.eu}
\affiliation{%
  \institution{Europeana Foundation}
  \city{The Hague}
  \country{Netherlands}
}

\author{Ellen Van Keer}
\email{ellen.vankeer@meemoo.be}
\affiliation{%
  \institution{Meemoo}
  \city{Ghent}
  \country{Belgium}
}

\author{Mahendra Mahey}
\email{mmahey@tlu.ee}
\affiliation{%
  \institution{Tallinn University}
  \city{Tallinn}
  \country{Estonia}
}

\author{Olga Holownia}
\email{olga@netpreserve.org}
\affiliation{%
  \institution{International Internet Preservation Consortium}
  \country{United States of America}
}

\author{Milena Dobreva}
\orcid{0000-0002-2579-7541}
\email{milena.dobreva@gmail.com}
\affiliation{%
  \institution{Institute of Mathematics and Informatics, Bulgarian Academy of Sciences}
  \city{Sofia}
  \country{Bulgaria}
}

\renewcommand{\shortauthors}{Candela and Gabriëls, et al.}

\begin{abstract}
  Large-scale digitization in Galleries, Libraries, Archives and Museums (GLAM) created the conditions for providing access to collections as data. It opened new opportunities to explore, use and reuse digital collections. Strong proponents of collections as data are the Innovation Labs which provided numerous examples of publishing datasets under open licenses in order to reuse digital content in novel and creative ways. 
  Within the current transition to the emerging data spaces, clouds for cultural heritage and open science, the need to identify practices which support more GLAM institutions to offer datasets becomes a priority, especially within the smaller and medium-sized institutions.

  This paper answers the need to support GLAM institutions in facilitating the transition into publishing their digital content and to introduce collections as data services; this will also help their future efficient contribution to data spaces and cultural heritage clouds. It offers a checklist that can be used for both creating and evaluating digital collections suitable for computational use. The main contributions of this paper are i) a methodology for devising a checklist to create and assess digital collections for computational use; ii) a checklist to create and assess digital collections suitable for use with computational methods; iii) the assessment of the checklist against the practice of institutions innovating in the Collections as data field; and iv) the results obtained after the application and recommendations for the use of the checklist in GLAM institutions.

\end{abstract}



\maketitle

\section{Introduction}

During the past few decades Galleries, Libraries, Archives and Museums (GLAM) have provided access to their collections and materials in digital format. Organisations have been exploring the benefits of adopting the concept of \textit{Labs} to publish under open licenses in order to reuse the digital collections in innovative and creative ways~ \cite{open_glam_lab}. Advances in technology have paved the way to publish digital collections suitable for computational use known as Collections as data \cite{padilla_thomas_2019_3152935}. Furthermore, with the emergence of initiatives such as the common European Data Space for Cultural Heritage\footnote{\url{https://pro.europeana.eu/page/common-european-data-space-for-cultural-heritage}} and the European Cultural Heritage Cloud,\footnote{\url{https://research-and-innovation.ec.europa.eu/research-area/social-sciences-and-humanities/cultural-heritage-and-cultural-and-creative-industries-ccis/cultural-heritage-cloud_en}} the need is even more urgent to incorporate Collections as data activities into the day-to-day operations of cultural heritage institutions in combination with building the necessary capacities to proactively contribute to such initiatives. 

Many GLAM organisations provide their contents for computational use in several ways. For instance, the Data Foundry at the National Library of Scotland provides metadata and digitised collections using a CC0 license.\footnote{\url{https://data.nls.uk/data/}} The Library of Congress provides access to information about historic newspapers and selected digitised newspaper pages as JSON, Linked Data and bulk data.\footnote{\url{https://chroniclingamerica.loc.gov/about/api/}} The Bibliothèque nationale du Luxembourg provides access to a  newspapers dataset with rich metadata using international XML standards such as Metadata Encoding and Transmission Standard (METS) and Analyzed Layout and Text Object (ALTO).\footnote{\url{https://data.bnl.lu/data/historical-newspapers/}} These initiatives can encourage other GLAM organisations to publish their collections suitable for computational use by following best practices and guidelines. However, as there is a wide diversity of approaches for publishing digital collections, organisations need proper assistance in selecting the best approach suited to their goals and, at the same time, considering researchers' and other reusers' needs. Several aspects might be considered in terms of how datasets are made available, including metadata formats (e.g., MARCXML, Dublin Core, JSON, etc.), data cleaning, licensing or documentation about the datasets.

This paper aims to define a checklist that can be used for both creating and evaluating digital collections suitable for computational use that are published by relevant institutions in the GLAM sectors. This approach provides an easy-to-apply method to encourage small and medium-size organisations to publish their digital collections as Collections as data. The main contributions of this paper are: i) a checklist to create and assess datasets suitable for use with computational methods; ii) the application of the checklist; and iii) the results obtained after applying it.

The paper is organised as follows: after a brief description of state of the art in Section~\ref{sec:sota}, Section~\ref{sec:framework} describes the methodology used to build the checklist. The application of the methodology and results are shown in Section~\ref{sec:application}. The paper concludes with an overview of the methodology and future work.

\section{Related Work}
\label{sec:sota}

The use of Artificial Intelligence and Machine Learning in the GLAM sectors has become a relevant topic aiming at applying new methods to the rich digital collections made available by the organisations \cite{loc_ia, padilla_oclc, padilla_thomas_2019_3152935, DBLP:conf/teachml/StrienBMT21}. In this sense, new initiatives on advancing the use of Artificial Intelligence have emerged such as Artificial Intelligence for Libraries, Archives and Museums (AI4LAM)\footnote{\url{http://www.ai4lam.org/}} and NewsEye.\footnote{\url{https://www.newseye.eu}} Several aspects regarding data quality and transparency in terms of how the data is available for the public (e.g., license, format, access, etc.) have become crucial elements for researchers willing to reuse the contents \cite{doi:10.1177/01655515211060530}. Many organisations such as the Bibliothèque nationale de France, the British Library and the Rijksmuseum focus on the application of new and advanced technologies to their digital materials \cite{bnf-data, DBLP:journals/eswa/DobbsR22, bl-opendata}. In addition, organisations have explored the benefits and challenges to use Application Programming Interfaces (APIs) in order to make available their digital collections as well as advanced vocabularies to describe the metadata \cite{harvard,moma, DBLP:journals/semweb/KohoILTTH21,metadata_oclc}. Moreover, features such as data cleaning and enrichment, the use of expressive controlled vocabularies instead of traditional metadata formats, using advanced and widespread APIs and the use of common and known open licenses have become crucial to facilitate the reuse of the contents. These technological innovations are relevant to the effors in building a data space for cultural heritage and need to meet the needs of different types of users \cite{ DBLP/conf/icadl/icadl2022}.

Despite all these efforts, there is still room for improvement regarding the publication of digital collections suitable for computational use \cite{doi:10.1177/01655515211060530}. Adopting these new trends from scratch might be difficult for organisations due to several reasons, e.g., the absence of dedicated personnel, a limited budget or the lack of advanced technical skills. 

In this context, a checklist provides a powerful tool as it presents a list of tasks, activities, and behaviors that need to be followed to achieve a systematic result. Checklists can be useful to help organisations to avoid common mistakes and to adopt best practices. In this way, the creation of checklists have emerged as an innovative method to provide best practices and guidelines. Several initiatives have tackled the definition and creation of checklists in the past in other domains, for instance for the improvement of the reliability of artificial intelligence systems in terms of the life cycle \cite{DBLP:conf/bigcomp/HanC22} and the evaluation of software process line approaches \cite{DBLP:journals/infsof/AghGP22}. Here, a checklist publication workflow was proposed including aspects such as source data management, reproducible data transformation, version control, data documentation and publication \cite{DBLP:journals/biodb/ReyserhoveDOASD20}. Other initiatives include a checklist for developing a machine learning project based on cultural heritage data \cite{DBLP:journals/corr/abs-2207-02960} and a checklist for a Data Management Plan made available by the Digital Curation Centre \cite{ddc-checklist}.

Regarding Collections as data, previous work has proposed a methodology to select datasets for computationally driven research applied to Spanish text corpora in order to encourage Spanish and Latin American institutions to publish machine-actionable collections \cite{doi:10.1177/01655515211060530}. A compilation of actions that can be done to stimulate conversation, and to encourage and generate ideas and new possibilities concerning the publication of digital collections suitable for computational use was recently published \cite{padilla_thomas_2019_3066237}. 

The use of advanced technologies such as Artificial Intelligence in combination with rich data made available by GLAM organisations raised important ethical issues \cite{romein_c_annemieke_2022_7267245, ethical}. These include, for example, control over the data, including terms of service requirements, the subsistence of the organisation sharing the data, the anonymous release of data and the threat of potential re-identification, and awareness of potential uses of the data. While clearer guidelines and better coordination are needed \cite{padilla_thomas_2019_3066237, ethical}, libraries and universities are in the position of playing a crucial role in education concerning unknown and future ethical issues. 

These efforts provide an extensive demonstration of how to make available digital collections suitable for computational use, giving particular attention to data quality, planning and experimentation. Nevertheless, to our best knowledge, none of the work to date provides an easy-to-follow and robust checklist to publish Collections as data in GLAM institutions. This checklist intends to encourage small- and medium-sized institutions to adopt the Collection as Data principle in their daily workflows following best practices and guidelines.

\section{A checklist to publish Collections as data in GLAM institutions}
\label{sec:framework}

Making available digital collections suitable for computational use is a complex process. Examples in the literature follow different approaches, making it difficult to adopt and standardise the process. In this sense, institutions may face challenges when addressing the adoption of Collections as data due to the lack of expertise, guidelines and best practices. This section introduces the methodology to create an easy to follow checklist to publish collections as data in the GLAM sector.

The checklist was constructed in four stages: i) relevant aspects for publishing digital collections suitable for computational use were identified based on existing implementations of the Collections as data principle and on a literature review; ii) potential issues and needs regarding how to make collections available as data were gathered from practitioners and researchers in GLAM and research institutions; iii) the checklist was built by synthesising the literature results and the issues and needs obtained in the previous steps; and iv) the checklist was tested and applied both as a tool for assessing a selection of datasets made available by GLAM institutions as proof of concept and as a supporting tool for creating collections as data.

The checklist proposed in this work is intended to encourage GLAM institutions to adopt Collections as data as a concept in their daily workflows. In addition, it could be distributed as additional and transparent information in the datasets for potential reusers and researchers.

\subsection{Previous works based on data published by GLAM}
The first step is based on a literature review encompassing existing work on publishing checklists in different domains and data management plans, institutional reports from GLAM organisations about digital collection publication for the public, and projects based on the reuse of the digital collections with innovative and creative approaches. In addition, recent research articles were searched in repositories (e.g., ACM Digital Library and dblp) about the impact and reuse of digital collections in GLAM institutions. Appendix \ref{app:review} shows the list of studies included in the review. The items were classified into five categories as shown in Table \ref{tab:categories}.

\begin{table}[h]
  \caption{Literature review to create the checklist classified into categories.}
  \label{tab:categories}
  \scalebox{0.9}{
  \begin{tabular}{cl}
    \toprule
    Category&References\\
    \midrule
    Best practices & \cite{padilla_thomas_2019_3152935,loc_ia, padilla_thomas_2019_3066237, sherratt_tim_2019_3551405, DBLP:journals/jis/RomeroSES22, open_glam_lab,metadata_oclc, wikibase_oclc,padilla_oclc,DBLP:journals/cacm/GebruMVVWDC21,lc1,lc3,romein_c_annemieke_2022_7267245,ethical}\\
    \hline
    Data quality & \cite{doi:10.1177/01655515211060530, DBLP:journals/jis/RomeroECS22, DBLP:conf/datech/Kiraly19,DBLP:conf/icaart/StrienBAHMC20}\\
    \hline
    Checklist definition & \cite{DBLP:journals/infsof/AghGP22, ddc-checklist,DBLP:journals/corr/abs-2207-02960, DBLP:conf/bigcomp/HanC22, DBLP:journals/biodb/ReyserhoveDOASD20}\\
    \hline
    Strategy \& data plan & \cite{bl-opendata,nls_data_plan,bnl_data_plan, bnf-data, austra-data, europeana_data, dpc_strategy, rluk, liber,nls-ai}\\
    \hline
    Examples \& experiments & \cite{DBLP:journals/eswa/DobbsR22, DBLP:journals/semweb/DijkshoornJAOSW18, moma, harvard, DBLP:journals/semweb/KohoILTTH21,DBLP:journals/jasis/RomeroC22,DBLP:journals/corr/abs-2005-01583,data-bne,data-bl,data-bnl,data-foundry, aus-data,DBLP:journals/semweb/RomeroECS18,lc2,lc4,kbr}\\
    
  \bottomrule
\end{tabular}
}
\end{table}

The items classified as \emph{best practices} include literature regarding guidelines to adopt Collections as data from different projects and authors, as well as recent approaches to publishing and reusing machine-actionable datasets. \emph{Data quality} encompasses research articles discussing the assessment of datasets with various methods as well as considering different types of contents such as text and metadata. \emph{Checklist definition} includes methodologies for creating checklists and examples applied to several domains. \emph{Strategy and data plan} describes reports made available by large institutions and initiatives regarding digital change. The \emph{examples and experiments} category introduces several examples of datasets and how they can be accessed and reused in innovative and creative ways.

\subsection{Identifying issues and information needs when implementing the Collections as data principle}
\label{sec:survey}

We conducted an observational study regarding the knowledge about and uptake of the Collections as data principle in GLAM institutions using an online survey during the period 10 to 30 October 2022. Participation was voluntary. We invited participants to provide the name of their institution and contact information while leaving open the option of anonymity. Consent was obtained from all respondents to include the survey results anonymously. 

A first core set of questions aimed to understand the respondents' existing experience with Collections as data, including the issues encountered in the early implementation phases, and collect examples of datasets already published. A second core set of questions was included to identify to what extent the respondents felt sufficiently informed when starting to implement the Collections as data principle and to understand their information needs. Table \ref{tab:survey} shows the questions used in the form sent to the participants.

\begin{table}
  \caption{Survey employed to retrieve information regarding the publication of Collections as data in GLAM institutions.}
  \label{tab:survey}
  \scalebox{0.85}{
  \begin{tabular}{lp{7cm}p{5cm}}
    \toprule
    Category & Question & Type\\
    \midrule
    Introduction & Goal of the survey & -\\
    \hline
    Contact Information & Institution, email, name, etc. & Text\\
    \hline
    \multirow{3}{*}{Experience with Collections as data}
    & What is the level of your experience with preparing Collections as data? & Scale 1-5 (1 = no experience; 5 = we have datasets ready and are confident that we know what to do)\\
    & Feel free to include (a) link(s) to your collection data sets here& Text\\
    & What were the main issues that you encountered when starting to prepare Collections as data? & Text\\

    \hline
    \multirow{3}{*}{Learning to prepare Collections as data} 
    & How well-informed do you feel / did you feel when starting to move towards Collections as data? & Scale 1-5 (1= not well-informed at all; 5 = very well-informed)\\
    & Main sources of information are / were (include as many as you wish) & Text\\
    & What information would you like to have / have liked to have had when starting to work towards Collections as data? What knowledge would have made it easier? & Text\\
    \hline
    Summary & Acknowledgement and contact & -\\
  \bottomrule
\end{tabular}
}
\end{table}

\begin{figure*}
\includegraphics[scale=0.5]{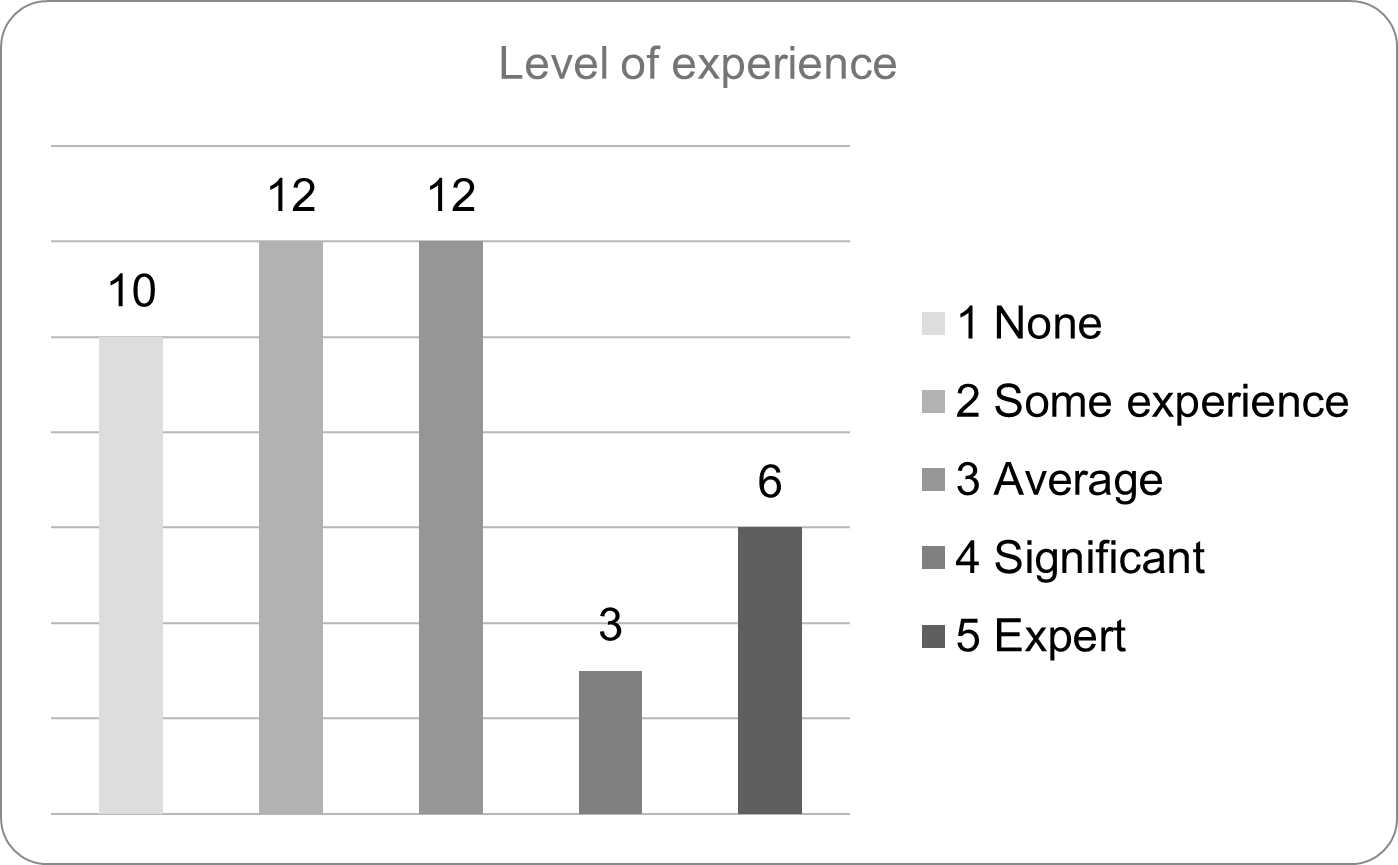}
\caption{Survey results "What is the level of your experience with preparing Collections as data?"}
\label{fig:surveygraph1}
\end{figure*}

\begin{figure*}
\includegraphics[scale=0.5]{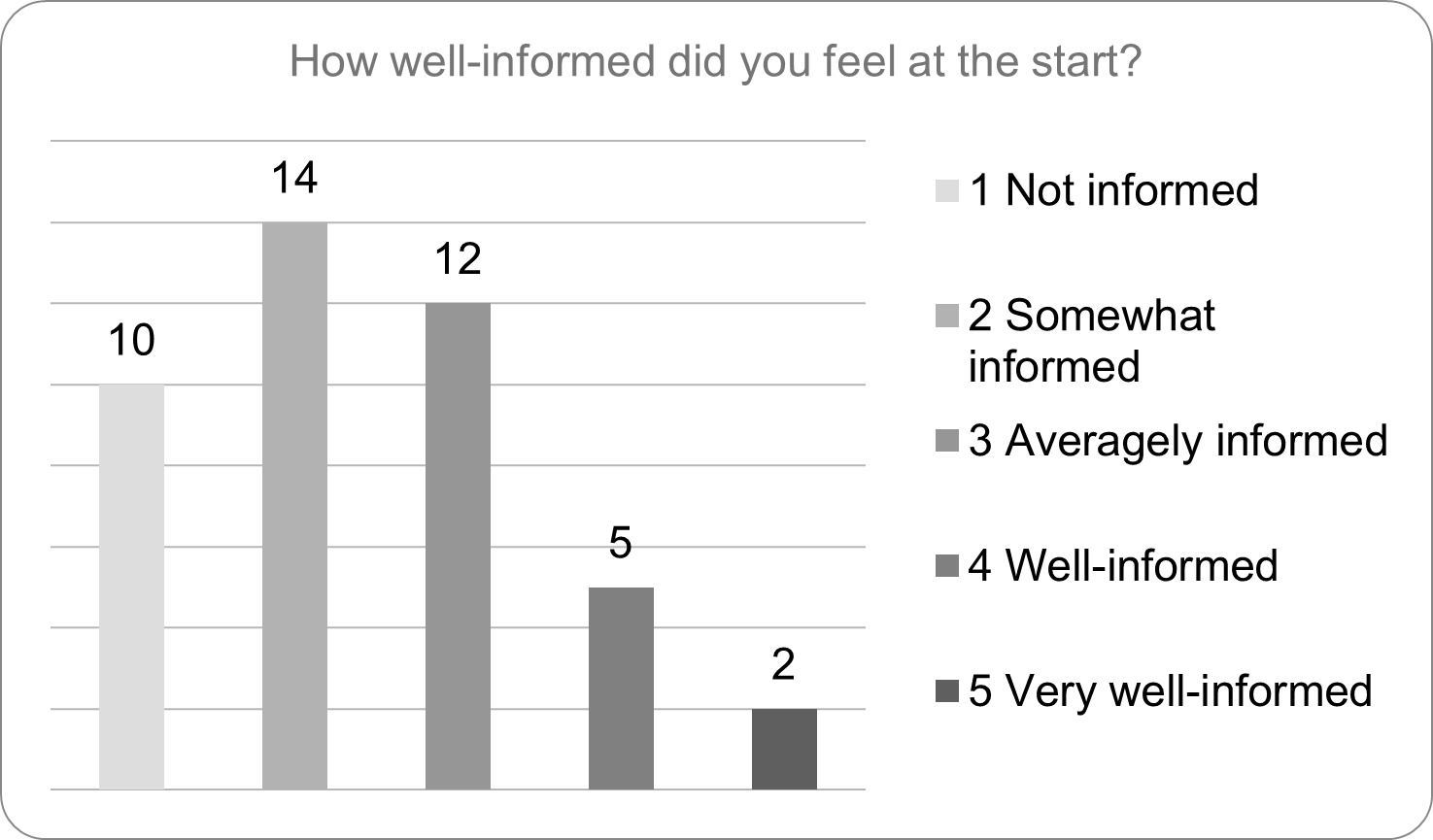}
\caption{Survey results "How well-informed do you feel/did you feel when starting to move towards Collections as data?"}
\label{fig:surveygraph2}
\end{figure*}

The forty three unique responses came from GLAM and research institutions with a geographical spread across the USA (26) and Europe (14), complemented by one Asian and two fully anonymous contributions. Figure \ref{fig:surveygraph1} shows that over half of the respondents indicated a low level of experience with preparing collections as data and nine were significantly experienced or experts. Similarly, the majority of respondents felt ill-informed when starting work on collections as data, with only two feeling very well-informed (Figure \ref{fig:surveygraph2}).

\begin{figure*}
\includegraphics[scale=0.5]{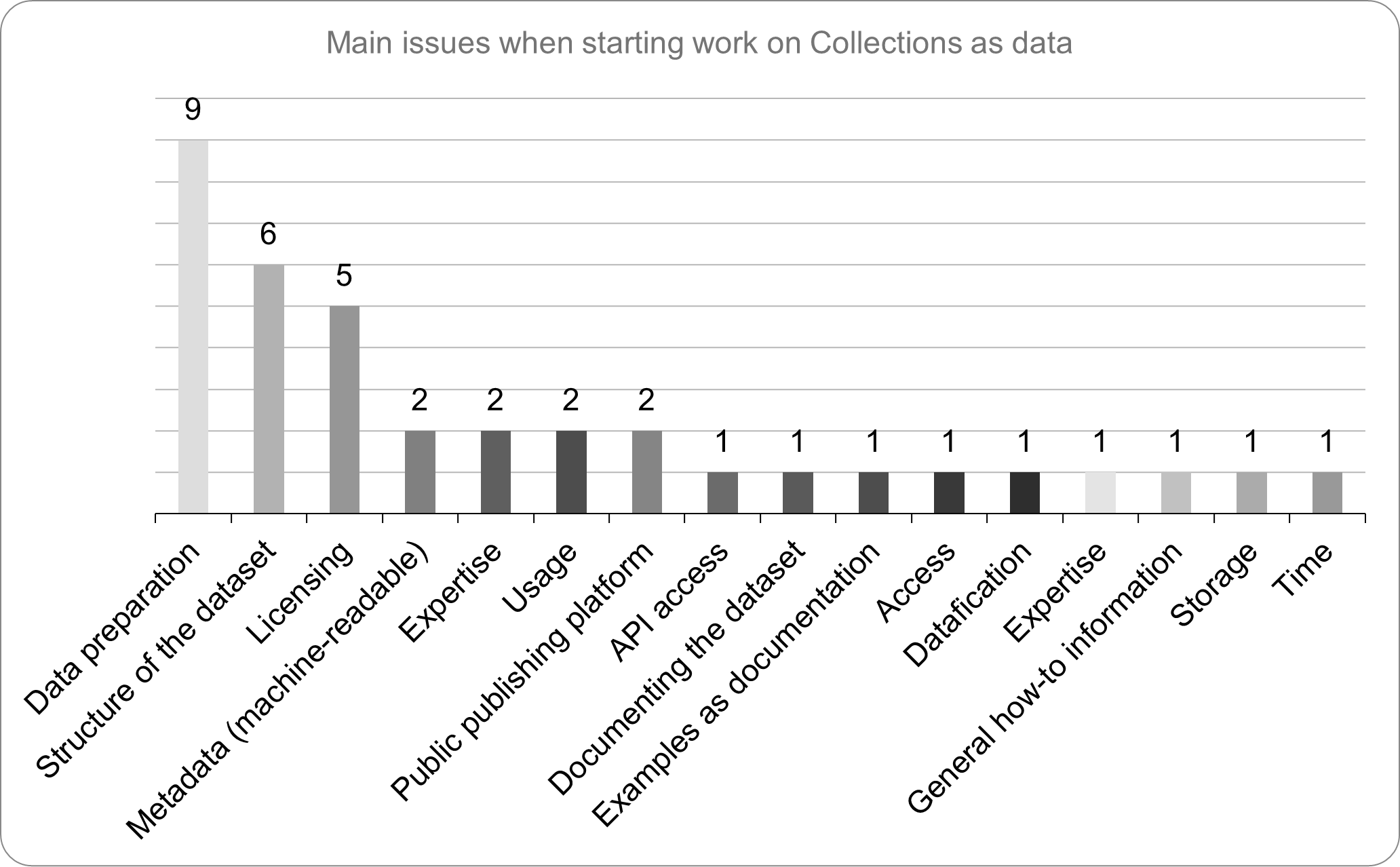}
\caption{Survey results "What were the main issues that you encountered when starting to prepare Collections as data?"}
\label{fig:surveygraph4}
\end{figure*}

The core issues encountered when creating collections suitable for computational use were those of data preparation and dataset structure as well as matters of licensing and usage restrictions (Figure \ref{fig:surveygraph4}). Data preparation is hampered by data quality issues, particularly regarding OCR data, but also because of incoherent data and inconsistencies, e.g. in the resources' descriptive metadata. Decisions on ontologies, vocabulary reconciliation, identifiers, overall structuring and packaging are all identified as obstacles when creating the dataset structure.   

\begin{figure*}
\includegraphics[scale=0.5]{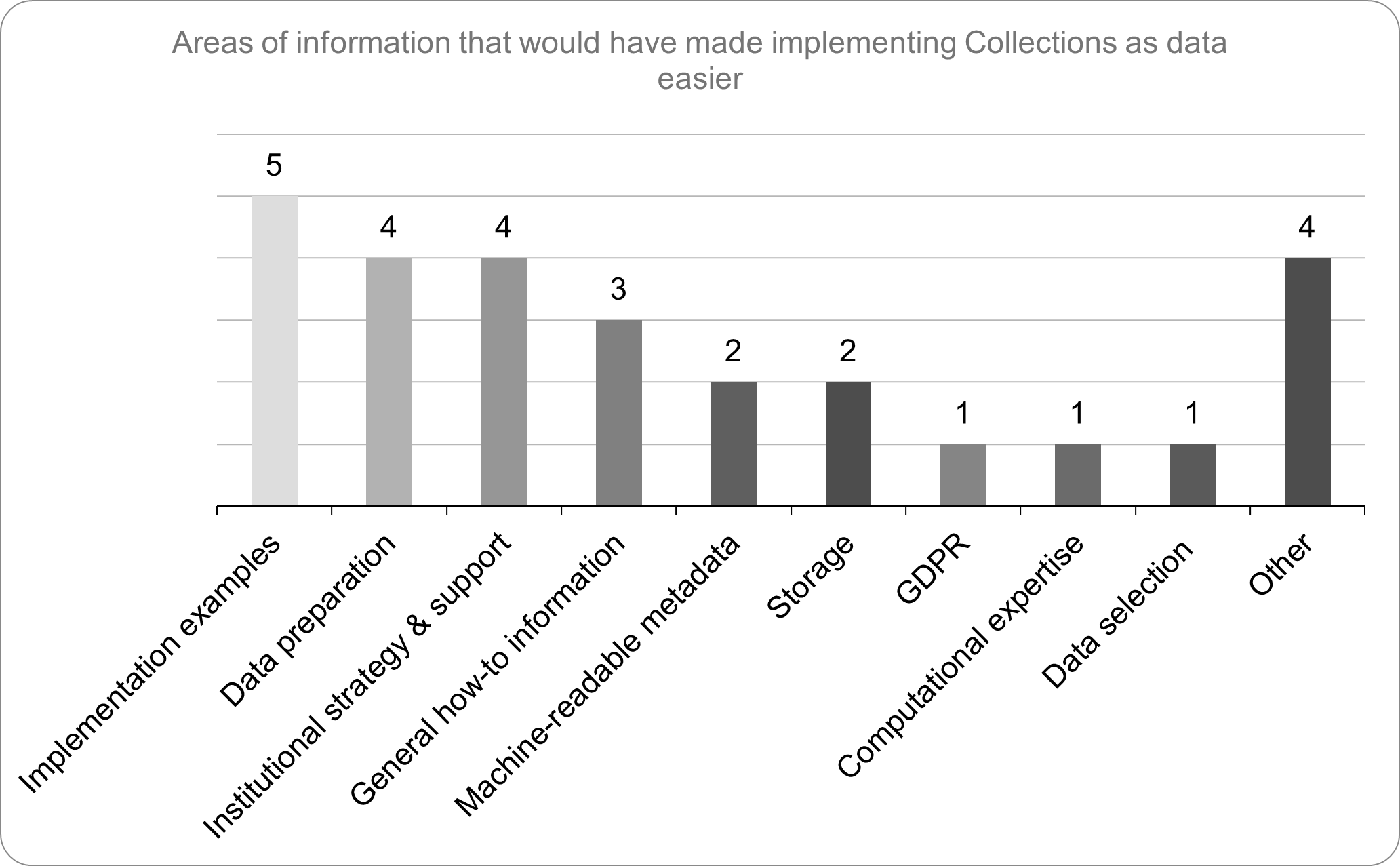}
\caption{Survey results "What information would you like to have/have liked to have had when starting to work towards Collections as data? What knowledge would have made it easier?"}
\label{fig:surveygraph3}
\end{figure*}

Figure \ref{fig:surveygraph3} reveals that institutions primarily name access to examples of implementation as well as both specific information on data preparation and general know-how about how to create collections as data as knowledge that would have simplified their uptake of the Collections as data principle. A register with collections as data projects and descriptions of the dataset creation processes would inspire and support institutions with no relevant experience in the initial implementation stages. Specific sought-after information on data preparation includes information on standards and best practices relating to file formats, metadata, data structure, and how to assess and (after selection) normalise the available data. General know-how should entail a user-friendly guide to tools, processes, decisions, and necessary policy choices and give insight in their implications on data modelling, data mapping and data reconciliation. On an organisational level, resources such as use cases showcasing the added value of collections as data could leverage strategic institutional support and encourage colleagues' and users' involvement. 

The survey further discloses a primary need for information and guidelines regarding the preparation of data and the structuring of datasets. Documented examples of datasets could significantly support GLAM institutions. Similarly, detailed accounts of the creation process for specific existing collections as data could inspire and support institutions in the decisions and actions to be taken when developing their own data for computational use. There is also a call for access to user-friendly guidelines and general know-how. The checklist explicitly intends to provide an easy-to-use tool in this context.

\subsection{The checklist}
\label{sec:checklist}
Based on the previous steps, a checklist to publish Collections as data was created as shown in Table \ref{tab:checklist}. A preliminary version was presented and discussed during an international webinar organised by the International GLAM Labs Community \cite{candela_gustavo_2022_7789480}.\footnote{\url{https://glamlabs.io/events/collections-data/}} An overview of each item is described below. 

\begin{table}
  \caption{Checklist to publish Collections as data in GLAM institutions.}
  \label{tab:checklist}
  \scalebox{0.95}{
  \begin{tabular}{rp{13cm}rr}
    \toprule
    Item & Description & Yes & No\\
    \midrule
    1 & Provide a clear license allowing reuse of the dataset without restrictions (e.g., CC0, CC BY) & &\\
    \hline
    2 & Provide a suggestion of how to cite your dataset & &\\
    \hline
    3 & Include documentation about the dataset & &\\
    \hline
    4 & Use a public platform to publish the dataset & &\\
    \hline
    5 & Share examples of use as additional documentation & &\\
    \hline
    6 & Give structure to the dataset & &\\
    \hline
    7 & Provide machine-readable metadata (about the dataset itself) & &\\
    \hline
    8 & Include your dataset in collaborative edition platforms & &\\
    \hline
    9 & Offer an API to access your repository & &\\
    \hline
    10 & Develop a portal page & &\\
    \hline
    11 & Add a terms of use & &\\
  \bottomrule
\end{tabular}
}
\end{table}

\subsubsection{Provide a clear license allowing reuse of the dataset without restrictions}
\label{subsec:license}

The adoption of licenses that allow reuse will strengthen and expand the role of GLAM institutions in innovative scholarly communication. The use of permissive licenses is crucial to ease an understanding of the reuse possibilities and to facilitate the reuse of the digital collections \cite{padilla_thomas_2019_3066237,DBLP:journals/jis/RomeroECS22}. Researchers expect a clear and reliable statement about the terms under which the dataset can be used. 

During the past few years, organisations have started to publish and promote the use of metadata and digital objects in their collections (or part of them) under open licenses \cite{nls_data_plan, doi:10.1177/01655515211060530, data-bnl, europeana_data, liber, bl-opendata}. While there are initiatives to develop national open licenses,\footnote{See, for example, \url{https://www.nationalarchives.gov.uk/doc/open-government-license/version/3/} and \url{https://www.data.gouv.fr/fr/pages/legal/licenses/}.} Creative Commons licenses are a popular and widely-used tool. Some examples of licenses, statements and tools used by GLAM institutions are the following:

\begin{itemize}
    \item The Creative Commons Public Domain Mark indicates that data is in the public domain. For instance, the Moving Image Archive published by the Data Foundry at the National Library of Scotland is published under this tool.\footnote{\url{https://data.nls.uk/data/metadata-collections/moving-image-archive/}}
    \item The Creative Commons Public Domain Dedication (CC0) removes copyright restrictions on the use of the content. For instance, the British Library and the Library of Congress provide a selection of datasets published under this tool.\footnote{See, for example, \url{https://www.bl.uk/collection-metadata/downloads} and \url{https://loc.gov/item/2018487899/}}
    \item CC BY data can be used when giving the appropriate credit to the source. For instance, the organisational data provided by the National Library of Scotland\footnote{\url{https://data.nls.uk/data/organisational-data/}} is published under a CC BY license. 
    \item National standards: other approaches are based on national licenses that describe how the data can be reused. For example, the Bibliothèque nationale de France made data available for the public on data.bnf.fr under the French Open license that enables the reuse and requires a attribution.
    \item Rights Statement No known copyright indicates that is likely to be free from copyright restrictions but the public domain cannot be entirely confirmed.
\end{itemize}

In addition, publication platforms such as GitHub and Zenodo allow users to select an appropriate license when publishing the contents. license information can be provided as textual information, including a link to the appropriate license\footnote{See, for example, \url{https://creativecommons.org/choose/}} or using metadata fields to describe copyright details such as the properties \texttt{dc:rights} and \texttt{dcterms:license} in the Dublin Core metadata schema.

Licensing the dataset must take into account the license of each of the resources contained in the dataset as these may vary. 

\subsubsection{Provide a suggestion of how to cite your dataset}
A suggestion for the citation promotes access and reusability of data as well as helps reusers to properly cite the dataset. Best practices recommend to include a preferred citation for the dataset \cite{padilla_thomas_2019_3066237}. 

A citation can be improved by using a permanent identifier to uniquely identify a resource such as a dataset~\cite{doi:10.1177/01655515211060530}. Digital Object Identifiers (DOI) are widely used by the community. For example, the datasets made available by the British Library and the National Library of Scotland provide a DOI as well as suggestions for citation. In fact, platforms such as Zenodo and DataCite provide a DOI for all published resources, including a citation in the most common citation formats such as BibTeX and APA.

Another practice is to describe the publication of a dataset in a research article that then can be used as a citation since journals provide citations in several formats. Several examples include the description of the transformation of a dataset into Linked Open Data (LOD) that have been made available as a research article~\cite{DBLP:journals/semweb/DijkshoornJAOSW18,DBLP:journals/semweb/KohoILTTH21}.

\subsubsection{Include documentation about the dataset}
Documentation is a key element to foster the reuse by the community \cite{padilla_thomas_2019_3066237}. Documentation may include details about the original sources as well as the cleaning and transformation principles and actions performed, information about how to access and use the dataset, or a description of the quality in terms of the content provided \cite{europeana_data}. 

The documentation can be provided in several ways such as a blog post, README files and tutorials. For example, Chronicling America provides information about the dataset by means of a dedicated website.\footnote{\url{https://chroniclingamerica.loc.gov/}} Other examples are based on the use of README files, as is the case for the British Library.\footnote{\url{https://www.bl.uk/collection-metadata/downloads}} 

\subsubsection{Use a public platform to publish the dataset}
Public platforms to make available datasets enable reusers to download the contents in bulk \cite{padilla_thomas_2019_3066237}. Some examples of free platforms are GitHub, Zenodo, Hugging Face and DataCite. However, some platforms may have limitations in terms of size for which paid services may be required. For example, the National Library of Scotland uses cloud storage services for their large datasets.\footnote{See, for example, \url{https://doi.org/10.34812/cg4r-dn40}}

These platforms provide additional features such as release management that can be useful to publish different versions of the same dataset \cite{romein_c_annemieke_2022_7267245}. 

\subsubsection{Share examples of use as additional documentation}
Examples of use of the contents provided by a digital collection are useful to inspire researchers \cite{padilla_thomas_2019_3066237, open_glam_lab}. 

In particular, a Lab environment within a GLAM organisation is the place where reusers are able to find examples and prototypes based on the digital collections that in many cases are made available under open licenses. For example, the KB Labs\footnote{\url{https://lab.kb.nl/}} from the National Library of the Netherlands provides a list of tools and the LC Labs from the Library of Congress include the experimental tool Newspaper Navigator that allows users to browse the images extracted from the digitised newspapers database Chronicling America \cite{DBLP:journals/corr/abs-2005-01583}.

In other cases, reproducible Jupyter Notebooks are used to introduce researchers to how to access and reuse the datasets. A Jupyter Notebook combines textual descriptions and code in the form of cells that can be run step by step. Some examples are the GLAM Workbench\footnote{\url{https://glam-workbench.net/}} and the GLAM Jupyter Notebooks from Biblioteca Virtual Miguel de Cervantes \cite{DBLP:journals/jis/RomeroSES22}.

Other approaches entail the publication of tutorials on platforms such as The Programming Historian~\cite{DBLP:conf/dihu/CrymbleMT12} and Library Carpentry \cite{city16083}, and research articles in journals describing how the dataset was created and reused.

\subsubsection{Give structure to the dataset}

A coherent internal distribution of a dataset is essential for researchers wishing to explore and query that dataset. Depending on the size and the type of contents, the structure will differ. Digital materials include a wide variety of content types, including images, maps, metadata, text, music and video amongst others.

There are some rules that will allow for a better understanding of the content provided by the dataset. One way to enhance this understanding is, for example, using self-describing folder names (e.g., text or images). Another approach could be based on the file format of the files provided (e.g., txt and xml). Each file included in the dataset may be named with the local identifier in the GLAM organisation. When having different formats for each resource (e.g., XML and JSON), a new root folder can be created clustering each of the formats.

For example, the Bibliothèque nationale du Luxembourg made available historical newspapers as open data using a zip file.\footnote{\url{https://data.bnl.lu/data/historical-newspapers/}} Each journal is included in a folder named with the title and the date. Each folder provides a set of folders according to different type of contents (images, pdf, text, thumbnails), the complete pdf and a xml file. Other approaches are based on metadata and provide a set of documents with different formats (e.g., Dublin Core and MARC) such as the Moving Image Archive.

More advanced initiatives such as BagIt File Packaging Format \cite{DBLP:journals/rfc/rfc8493}, describes a set of hierarchical file layout conventions for storage and transfer of arbitrary digital content. 

When providing large-size images, which is often the result of a digitization process, it can be interesting to provide reduced-size thumbnails based on the original images to be able to visualise them easier and faster. One additional aspect to consider is the cleaning of the data before publication. For example, sometimes post-correction OCR data is included in the case of digitization datasets, or metadata collections may require cleaning to remove unnecessary metadata fields.

\subsubsection{Provide machine-readable metadata}
There is a wide variety of forms and formats to make metadata about digital resources (e.g., a dataset) available. The use of interoperable machine-readable metadata enhances discoverability and use since the data is readily processed by a computer \cite{w3cData}. Some examples of vocabularies to provide metadata are MARC, Dublin Core, Vocabulary of Interlinked Datasets (VoID) \cite{void} and Data Catalog Vocabulary (DCAT) \cite{dcat}. Other initiatives are based on Resource Description Framework (RDF) and schema.org. For example, the machine readable metadata description using the vocabulary DCAT for the dataset National Bibliography of Scotland published by the Data Foundry is shown in Listing \ref{code1}.

\begin{lstlisting}[basicstyle=\small,language=C,frame=single,caption=Machine-readable metadata description using the vocabulary DCAT for the dataset National Bibliography of Scotland published by the Data Foundry,label=code1]

@prefix dcat: <http://www.w3.org/ns/dcat#> .
@prefix dct: <http://purl.org/dc/terms/> .

<https://doi.org/10.34812/7cda-ep21> a dcat:Distribution ;
  dcat:downloadURL <https://nlsfoundry.s3.amazonaws.com/metadata/nls-nbs-v2.zip> ;
  dct:license <https://creativecommons.org/licenses/by/4.0/> ;
  dcat:mediaType <https://www.iana.org/assignments/media-types/text/xml> ;
  dcat:compressFormat <http://www.iana.org/assignments/media-types/application/zip>
.
\end{lstlisting}

\subsubsection{Include your dataset in collaborative edition platforms}
Collaborative edition platforms have become increasingly relevant in the GLAM context to create links and enrich their collections \cite{padilla_thomas_2019_3066237, wikibase_oclc, DBLP:journals/jis/RomeroECS22}. Crowdsourcing approaches enable the community to contribute to the content in a collaborative environment. 

Wikidata, for example, enables the creation of resources known as entities, adding properties to describe the entities. The edition is performed using an easy and accessible web interface. For example, the section dedicated to computational access to digital collections at the International GLAM Labs Community website includes a selection of Jupyter Notebooks projects made available by relevant institutions that have been published in Wikidata \cite{glamlabs}.\footnote{See, for example, the resource \url{https://www.wikidata.org/wiki/Q111421205}} Wikidata provides a public API to access the data, enabling users to retrieve the contents. Table \ref{tab:wikidataprop} shows an overview of Wikidata properties that can be useful to describe datasets.

\begin{table}
  \caption{Overview of Wikidata properties to describe a dataset as an entity.}
  \label{tab:wikidataprop}
  \scalebox{0.95}{
  \begin{tabular}{lp{2cm}p{9cm}}
    \toprule
    Property & Identifier & Description\\
    \midrule
    full work available at URL & P953 & full work available at URL\\
    \hline
    instance of & P31 & that class of which this subject is a particular example and member. For example, newspaper archive and data set.\\
    \hline
    language of work or name & P407 & language associated with this creative work\\
    \hline
    owned by & P127 & owner of the subject\\
    \hline
    publication date & P577 & date or point in time when a work was first published or released\\
    \hline
    title & P1476 & published name of a work\\
  \bottomrule
\end{tabular}
}
\end{table}

\subsubsection{Offer an API to access your repository}
The use of an API to make available the dataset is a key element to foster reuse \cite{padilla_thomas_2019_3066237}. APIs allows systems to communicate and to access and retrieve the entire dataset. In some cases, only a portion of the dataset may be retrieved for analysis using the API.

The use of an API to publish the digital contents may require additional features to be considered. For instance, when using IIIF, each resource should include a manifest.xml describing the contents of this resource. For LOD, the adoption of URL patterns for the resources (e.g., author/id or author/name) is required as well as an analysis of how the data will be modelled (e.g, classes used and number of properties) according to the controlled vocabularies used to describe the metadata. 

Table \ref{tab:api} introduces an overview of digital collections made available by institutions using a wide variety of APIs.

\begin{table}
  \caption{Overview of digital collections made available by relevant institutions using a wide variety of APIs.}
  \label{tab:api}
  \scalebox{0.95}{
  \begin{tabular}{lp{2cm}p{8cm}}
    \toprule
    Dataset & API & URL\\
    \midrule
    Biblioteca Virtual Miguel de Cervantes & Linked Data & \url{https://data.cervantesvirtual.com/sparql}\\
    \hline
    Bibliothèque nationale du Luxembourg & OAI-PMH & \url{https://data.bnl.lu/apis/oai-pmh/}\\
    \hline
    Chronicling America & JSON & \url{https://chroniclingamerica.loc.gov/about/api}\\
    \hline
    Deutsche Nationalbibliothek & OAI-PMH & \url{https://www.dnb.de/EN/oai}\\ 
    \hline
    Harvard & IIIF & \url{https://iiif.harvard.edu/about-iiif/}\\
    \hline
    Harvard Art Museums & JSON &
    \url{https://harvardartmuseums.org/collections/api}\\
    \hline
    Library of Congres & JSON/YAML & \url{https://www.loc.gov/apis/}\\
    \hline
    Museum of Modern Art  & JSON or XML & \url{https://api.moma.org/}\\
    \hline
    Victoria and Albert Museum & IIIF & \url{https://developers.vam.ac.uk/guide/v2/images/iiif.html}\\
    \hline
    WarSampo & Linked Data & \url{https://www.ldf.fi/}\\
  \bottomrule
\end{tabular}
}
\end{table}

\subsubsection{Develop a portal page}
Using a portal page for the dataset enhances the visibility and facilitates additional information about reusing the data \cite{padilla_thomas_2019_3066237}. This information may include references to links for the dataset, visualisations, awards received, contact information, etc. For example, the dataset Chronicling America includes a dedicated website to access the contents but also to understand how the API can be used and to provide information about the original sources and license.

In addition, platforms such as GitHub provide free services to publish websites that are stored as a code repository and enables the use of several themes.\footnote{\url{https://pages.github.com/}}

\subsubsection{Add a terms of use}
Best practices show the importance of adding terms of use describing the conditions of use for the data \cite{padilla_thomas_2019_3066237}. The content can be provided as an additional section on a portal page or as a text document. 

For example, the British Library EThOS dataset includes a terms of use section that details copyright, liability and access statements.\footnote{\url{https://ethos.bl.uk/ViewTerms.do}} Other examples describe additional aspects such as how to report content as inappropriate in situations where people's rights are violated.\footnote{\url{https://loar.kb.dk/assets/pdfs/LOAR_TermsofService_v3_KB.pdf}}

\section{Application of the checklist}
\label{sec:application}
The checklist is intended as a tool for institutions to start implementing the Collections as data principle by giving a list of actions that can be performed so as to make collection data ready for computational use and reuse. Whilst not all items on the checklist must be executed to present collections as data, they give a clear direction when deciding on which action to prioritise and which to defer to a later stage or to consider as unfitting to the specific collection. The checklist also serves as a tool for assessing existing collections as data, indicating the level of readiness for potential computational use. 

This section provides cases for both of the above uses: it presents the results of the application of the checklist to assess a selection of datasets made available by relevant GLAM institutions as well as describes cases where the checklist provided handles to implement the Collection as data principle in an institutional context.

\subsection{The checklist as a tool for assessing GLAM collections as data}

A selection of datasets made available by a wide variety of GLAM institutions in terms of size has been assessed against the checklist (see Table \ref{tab:datasets}). The institutions and datasets are listed below.

\begin{itemize}
    \item The British Library has a number of data services available to support different use cases, for example the content showcased on data.bl.uk and hosted on the open access British Library Research Repository.\footnote{\url{https://bl.iro.bl.uk/}} Some of these datasets will also have a corresponding Collection Guide.\footnote{See, for example, the example \url{https://www.bl.uk/collection-guides/digitised-printed-books}} 
    \item Data Foundry is the National Library of Scotland’s open data platform, which includes digitised datasets, metadata, spatial data and organisational data. For this example, we have chosen ‘Encyclopaedia Britannica’, the most-used dataset. This covers the first 8 editions (100 years) of the Encyclopaedia.\footnote{\url{https://doi.org/10.34812/cg4r-dn40}}
    \item The Library of Congress (LC) recently published \url{data.labs.loc.gov}, as an experimental sandbox for sharing data packages compiled as part of LC Labs' Mellon Foundation-funded Computing Cultural Heritage in the Cloud (CCHC) initiative.\footnote{\url{https://labs.loc.gov/work/experiments/cchc/}} In this context, the Stereograph Card dataset consists of 39,526 stereograph card images from the 1850s through 1924, a subset of what was available online in the collection in the catalog in August 2022. 
    \item The Royal Danish Library made available an API for its digitised collection as a result of a newspaper digitization project running from 2014 to 2017. The construction of the API has been a way to experiment with the OpenAPI standard.\footnote{\url{https://swagger.io/specification/}} 
    \item Art In Flanders (AIF) is a dataset supported by Meemoo that includes more than 20.000 images of objects from Flemish museums and cultural institutions, comprising paintings, sculptures, archaeological artefacts, design objects, and more. Digital reproductions and descriptive metadata are being made available through the artinflanders.be platform. 
    \item Miguel de Cervantes Virtual Library (BVMC) made available its main catalogue as Linked Open Data (LOD) using Resource, Description and Access (RDA) as its main vocabulary \cite{DBLP:journals/semweb/RomeroECS18}. 
\end{itemize}

Table \ref{tab:results} shows the results obtained after the assessment in terms of the items provided by the checklist introduced in Section \ref{sec:framework}.

\begin{table}[h]
  \caption{Overview of the datasets and organisations used for the assessment.}
  \label{tab:datasets}
  \scalebox{0.70}{
  \begin{tabular}{p{3.5cm}lp{6cm}p{3cm}}
    \toprule
    Organisation & URL & Dataset description & license\\
    \midrule
    British Library & \url{https://www.bl.uk/collection-guides/digitised-printed-books} & 
Digitised printed books (18th-19th century) & Public Domain Mark\\
    \hline
    Library of Congress & \url{https://data.labs.loc.gov/stereographs/} & 
39,526 stereograph card images from the 1850s through 1924 & Library's statement\\
    \hline
    Miguel de Cervantes Virtual Library &
    \url{https://data.cervantesvirtual.com/datos-enlazados} & Main catalogue as LOD & Creative Commons CC0 1.0 Universal Public Domain Dedication\\
    \hline
    Meemoo &
    \url{https://artinflanders.be/en} & 20.000 images of objects from Flemish museums and cultural institutions & Public Domain Mark and in-copyright\\
    \hline
    National Library of Scotland's Data Foundry & \url{https://doi.org/10.34812/cg4r-dn40} & 
Encyclopaedia Britannica & Public Domain Mark
 \\
    \hline
    Royal Danish Library & \url{https://www2.statsbiblioteket.dk/mediestream/avis} & Digitised newspaper collection & Public Domain Mark\\
  \bottomrule
\end{tabular}
}
\end{table}

\begin{table}[h]
  \caption{Overview of the results obtained when evaluating the checklist against a list of datasets made available by relevant GLAM institutions.}
  \label{tab:results}
  \scalebox{0.95}{
  \begin{tabular}{lrrrrrrrrrrr}
    \toprule
    Organisation & 1 & 2 & 3 & 4 & 5 & 6 & 7 & 8 & 9 & 10 & 11\\
    \midrule
    British Library & \ding{51} & \ding{51} & \ding{51} & \ding{51} & \ding{51} & \ding{51} & \ding{51} & \ding{51} & \ding{51} & \ding{51} & \ding{51}\\
    \hline
    Library of Congress & \ding{51} & \ding{51} & \ding{51} & \ding{51} & \ding{51} & \ding{51} & \ding{51} & - & \ding{51} & \ding{51} & \ding{51}\\
    \hline
    Meemoo & \ding{51} & \ding{51} & \ding{51} & \ding{51} & - & \ding{51} & \ding{51} & - & - & \ding{51} & \ding{51}\\
    \hline
    Miguel de Cervantes Virtual Library & \ding{51} & \ding{51} & \ding{51} & \ding{51} & \ding{51} & \ding{51} & \ding{51} & \ding{51} & \ding{51} & \ding{51} & \ding{51}\\
    \hline
    National Library of Scotland & \ding{51} & \ding{51} & \ding{51} & \ding{51} & \ding{51} & \ding{51} & \ding{51} & - & - & \ding{51} & \ding{51}\\
    \hline
    Royal Danish Library & \ding{51} & - & \ding{51} & \ding{51} & - & \ding{51} & - & \ding{51} & \ding{51} & \ding{51} & \ding{51}\\
  \bottomrule
\end{tabular}
}
\end{table}

\textbf{License.} All the datasets and platforms assessed provide a clear license. For example, the British Library has a formal access and reuse process to identify if works are out of copyright or in copyright, the National Library of Scotland's Data Foundry provides the license for each dataset\footnote{See, for example, \url{https://data.nls.uk/data/digitised-collections/encyclopaedia-britannica/}} and the Library of Congress bases its reuse policies on its rights statement on the source collection. Table \ref{tab:datasets} introduces the licenses used in the dataset.

\textbf{Suggested citation.} In general, most of the datasets provide a persistent identifier such as a DOI. Other examples such as the National Library of Scotland's Data Foundry provide a suggested citation. The Library of Congress offers citations details for the source collections and the dataset creators and contributors. In other cases, a journal research article can be used to cite a dataset.

\textbf{Documentation about the dataset.} All the datasets provide dataset information and metadata as documentation in a wide diversity of manners (e.g., website, README) and granularity (e.g., collection and individual level). Other approaches are based on the use of machine-readable vocabularies based on RDF to describe the datasets such as the Vocabulary of Interlinked Datasets (VoID).\footnote{\url{http://vocab.deri.ie/void}} Table \ref{tab:documentation} shows an overview of the approaches followed by the institutions selected in this work.

\begin{table}[h]
  \caption{Overview of strategies to provide documentation about machine-actionable collections in GLAM institutions.}
  \label{tab:documentation}
  \scalebox{0.95}{
  \begin{tabular}{lp{8cm}}
    \toprule
    Organisation & Documentation strategies \\
    \midrule
    British Library & Website and datasheets\\
    \hline
    Library of Congress & Website, README, cover sheet, data processing plan\\
    \hline
    Meemoo & Website\\
    \hline
    Miguel de Cervantes Virtual Library & Website, research journal, VoID file\\
    \hline
    National Library of Scotland & Website\\
    \hline
    Royal Danish Library & Website\\
  \bottomrule
\end{tabular}
}
\end{table}

For example, the Library of Congress provides three levels of documentation for the datasets made available on data.labs.loc.gov by means on different files: i) a README file with a technical overview of how the data set was created (e.g., details of the dataset source collection, computational readiness and possible uses, dataset field descriptions and rights statement); ii) a data cover sheet file with a more substantive overview of the data and the collection from which it is derived (e.g., version information, background of collection, original format, reading room details, contact and metadata types); and iii) a data processing plan describing the goal of the experiment and a description of intended use, and data documentation regarding different aspects such as composition, provenance, compilation methods, preprocessing steps and potential risks to people and communities, amongst others.\footnote{\url{	https://blogs.loc.gov/thesignal/2022/12/announcing-lc-labs-data-sandbox-and-3-new-data-packages/}} 

In addition, the BL have explored innovative approaches such as Datasheets for Datasets by including a datasheet in the datasets, documenting its motivation, composition, collection process, recommended uses, etc. to facilitate better communication between dataset creators and dataset consumers.\footnote{\url{https://blogs.bl.uk/digital-scholarship/2022/04/making-british-library-collections-even-more-accessible.html}}

\textbf{Use of a public platform to publish the dataset.} All the datasets are available by means of public platforms. However, there are differences accross the institutions regarding the use of institutional and third party platforms. For example, the BL uses both institutional and third party platforms, including British Library Research Repository,\footnote{\url{https://bl.iro.bl.uk/}} Flickr, Wikimedia, Hugging Face, and secondary publishers, depending on the type/format of data. In the case of the other institutions, the datasets are available by means of an institutional website (e.g., Lab section and dedicated website) as is the case for Meemoo, BVMC and Data Foundry.

Other organisations have different approaches depending on the content provided. For example, the Library of Congress provides access to datasets that have been officially acquired in the Selected Datasets Collection.\footnote{\url{https://www.loc.gov/collections/selected-datasets/about-this-collection/}} For experimental or temporary datasets, access is provided on LC for Robots\footnote{\url{https://labs.loc.gov/lc-for-robots/}} or on data.labs.loc.gov which hosts datasets using cloud service providers.

\textbf{Share examples of use.} Many of the datasets assessed include examples of use as additional documentation to show how to reuse the contents. However, the approaches differ from one institutions to another. For example, the National Library of Scotland's Data Foundry provides examples based on reproducible Jupyter Notebooks and the project includes collaboration initiatives based on the reuse of the datasets.\footnote{See, for example, \url{https://data.nls.uk/projects/}} The BL shares examples of dataset reuse on its Digital Scholarship blog.\footnote{\url{https://blogs.bl.uk/digital-scholarship/}} The Library of Congress includes a section “Computational Readiness and possible uses” in the README files.\footnote{See, for example, \url{https://data.labs.loc.gov/stereographs/README.txt}}

\textbf{Give structure to the dataset.} While datasets are structured according to different requirements and contents, in general, datasets are structured with reuse and data management in mind. For example, the National Library of Scotland's Data Foundry provides the datasets as zip files including folders per file format that can be easily identified by potential reusers. The Library of Congress has explored different ways to create and communicate coherence in datasets structure. Some examples are including dataset field descriptions in the README files, metadata and manifests for scripted and API access, and providing sample data and guidance on each data package page for ways to download the OCRed text, documentation and metadata.

\textbf{Provide machine-readable metadata.} All the datasets provide machine-readable metadata to describe the digital collections based on Dublin Core and more advanced approaches based on controlled vocabularies. The metadata is provided in form of additional files (e.g., XML) or through an API.

\textbf{Include your dataset in collaborative edition platforms.} While many institutions already provide information about their datasets in Wikidata, they are also interested to develop further Wikidata opportunities. Table \ref{tab:wikidata} shows an overview of the Wikidata approaches in GLAM organisations. However, it is important to notice that in some cases the information included is not related to datasets but to other initiatives such as projects and notebooks. In addition, other approaches are based on Wikimedia approaches. For example, a subset of the BL dataset is currently on Wikimedia Commons,\footnote{\url{https://commons.wikimedia.org/wiki/Commons:British\_Library/Mechanical\_Curator\_collection}} which offers a useful introduction to the collection, including a Synoptic Index, as well as projects to georeference maps found in the texts.

\begin{table}[h]
  \caption{Overview of Wikidata links related to GLAM organisations.}
  \label{tab:wikidata}
  \scalebox{0.95}{
  \begin{tabular}{ll}
    \toprule
    Organisation & Wikidata link\\
    \midrule
    British Library & \url{https://www.wikidata.org/wiki/Wikidata:British\_Library}\\
    \hline
    National Library of Scotland & \url{https://www.wikidata.org/wiki/Q111411199}\\
    \hline
    Miguel de Cervantes Virtual Library & \url{https://www.wikidata.org/wiki/Q111396572}\\
  \bottomrule
\end{tabular}
}
\end{table}

\textbf{Offer an API to access your repository.} Most of the institutions have adopted APIs as a means to make available their collections based on different standards and tools. For example, the BVMC provides the datasets through a SPARQL public endpoint. Others provide a blend of this recommendation, using APIs to source collections and using manifests from data packages to gather those packages via a JSON/YAML API such as the Library of Congress.

However, some institutions have decided to provide the collections with simple, straightforward access through downloads to cater for those users whose technical skills are limited, such as students and artists.

\textbf{Develop a portal page.} All the institutions provide a data portal page including detailed information about the collections. Several examples included in this selection of organisations and datasets are the result of previous experimental data access points that have evolved to a new section such as Data.labs.loc.gov\footnote{https://data.labs.loc.gov/} and data.cervantesvirtual.com.

\textbf{Terms of use.} Terms of use are provided by the organisations. In some cases, the information includes contact details, for example the BL.\footnote{\url{https://bl.iro.bl.uk/terms}} In other examples, the information is provided only in the country's official language (e.g., Spanish).\footnote{See, for example, \url{https://data.cervantesvirtual.com/condiciones-de-uso/}} Table \ref{tab:terms} shows an overview of the terms of use provided by the institutions.

\begin{table}[h]
  \caption{Overview of terms of use provided by the organisations selected in this work.}
  \label{tab:terms}
  \scalebox{0.95}{
  \begin{tabular}{ll}
    \toprule
    Institution & URL\\
    \midrule
    British Library & \url{https://bl.iro.bl.uk/terms}\\
    \hline
    Miguel de Cervantes Digital Library & \url{https://data.cervantesvirtual.com/condiciones-de-uso/}\\
    \hline
    Data Foundry & \url{https://data.nls.uk/about/rights/}\\
    \hline
    Meemoo & \url{https://artinflanders.be/en}\\
  \bottomrule
\end{tabular}
}
\end{table}

\subsection{Providing a clear license: the case of meemoo and the Art in Flanders dataset}

Providing a clear license allowing reuse of the dataset with as few restrictions as possible is one of the more complex items on the checklist. Managing these rights over time only adds to this complexity, as is shown in the case of meemoo's Art in Flanders (AIF) dataset. Rights management has gained much attention in the heritage sector over the last few years, in relation to developments in computing (e.g. Linked Open Data) and copyright legislation (e.g. EU directive on copyright). In line with this, the need arose to review the rights labelling policy for the AIF platform. The ambition was twofold: i) to make the dataset as freely available for access and reuse as possible; and ii) to adopt more appropriate standard rights labels for communicating the rights information. In this sense, some issues were identified:
\begin{itemize}
    \item over the years inadequacies had slipped into the rights information on the AIF platform. To give an example: a painting that had fallen into the public domain because its creator died more than 70 years ago, but with a copyright waiver (CC0) in the metadata and a photo credit © name-photographer on the picture. 
    \item new copyright is claimed on digital surrogates of public domain works. The initiative wanted to get in line with article 14 of the EU directive that protects the public domain from surrogate rights. For so-called “2-dimensional” works (such as paintings) there is a clear consensus that no new copyright can be claimed on purely technical reproductions. However, for photographic reproductions of 3-dimensional objects, the situation is less straightforward as these may meet the threshold of originality because of the photographer’s personal choices in point of view, lighting etc. So copyright and copyright restrictions might still apply here. An additional complexity is that the terms of meemoo’s older photography contracts cannot be transposed into a standard open license, so more restrictive CC licenses are being used. 
    \item owners may impose restrictions on the use of reproductions made of works in their collections, even if they are in the public domain. Lacking more appropriate tools, these use conditions have been (improperly) translated into a restrictive CC license on the AIF platform.
\end{itemize}

The group of photographic reproductions of 3-dimensional artworks were particularly problematic, as the project was confronted with double-layered rights statuses. Two solutions were successively considered: 1) using separate rights labels: one for the rights status of the artwork and one for the rights status of the photo, and 2) using a single rights label that communicates the rights status and usage conditions for the resource as a whole. In parallel, the possibility of recontacting the photographers in question, and asking them to waive their rights, was considered. 

For images where the owner of the cultural objects imposes use restrictions on reproductions, it was decided to adopt a \textit{rightstatements.org} label. These labels have been specifically devised for heritage institutions to communicate rights information in a standardised way when they do not own the copyright and therefore using a copyright license is legally not possible. 

For the reproductions of two-dimensional works, it was proposed to use updated labels for the three main groups. Firstly, the majority are in the public domain and can be released with a public domain mark instead of a copyright waiver. These images are freely downloadable in high resolution and reusable for any purpose without any restrictions. Secondly, where collection owners have imposed use conditions, it was proposed to use the rights statement “contractual restrictions”. However, since this is not a legally binding tool, the user still needs to agree to user terms before downloading the images. In parallel, these contracts are currently being reviewed with the goal of minimizing and standardizing the user restrictions far as possible. Thirdly, when works are under copyright, images get the “in copyright” mark.

For the reproductions of 3-dimensional works, a one-label approach was chosen for a number of reasons. Firstly, a single rights label is more user-friendly. It requires less pre-existing knowledge and leaves less room for (mis)interpretation by the user than the multiple-label approach. So in a sense, this is also the more secure and controlled approach. Secondly, not all labels are entirely compatible. For instance, a CC BY license which allows reuse of an image of an artwork, but which is also under full copyright and cannot be used without permission of the rights owner. In these cases, it was proposed to use the most restrictive label. In this way, pictures of artworks that are in copyright are tagged as such, even when the photographer agrees to a more open license. Additionally, pictures in the public domain get their label from the license agreed upon with the photographer. 

In parallel, it was decided to include a maximum waiver of rights in (new) photography contracts. The process of clearing rights for older contracts is also in progress. 

In summary, the updated ‘open’ licensing policy on the AIF platform proposes 4 rights labels for the main categories of images:

\begin{itemize}
    \item Public Domain mark for images of 2D artworks in the public domain.
    \item CC0 for images of 3D artworks in the public domain.
    \item No-copyright - contractual restrictions for images of artworks in the public domain restricted by the collection owner (2D and 3D).
    \item In Copyright mark for images of artworks that are under copyright (2D and 3D).
\end{itemize}

The majority of images on AIF belong to the first two groups. In addition, providing access to the AIF dataset through an API is on the roadmap.

\subsection{The checklist as a tool for implementing the Collections as data principle at KU Leuven Libraries}
\label{sec:leuven}

As apparent from the survey results presented in Section \ref{sec:survey}, when working towards disclosing collections as data, GLAM institutions are looking for inspiration and general insight in how to approach the implementation of the Collections as data principle. The checklist can give guidance to the process, allowing to make informed decisions on what to focus on first. To give potential users of the checklist insight into this process, this section describes the case of KU Leuven Libraries'\footnote{https://bib.kuleuven.be/english} work on creating datasets for computational use.

\begin{table}[h]
  \caption{Overview of datasets created by KU Leuven Libraries for the BiblioTech 2023 Hackathon (13-23 March 2023). At the time of writing, the datasets are only accessible for KU Leuven staff and students.}
  \label{tab:leuvendatasets}
  \scalebox{0.95}{
  \begin{tabular}{p{4cm}p{6cm}l}
    \toprule
    Dataset & URL to physical collection & Hackathon team poster DOI\\
    \midrule
    Lecture Notes from the Old University of Leuven & \url{https://kuleuven.limo.libis.be/discovery/collectionDiscovery?vid=32KUL_KUL:KULeuven&inst=32KUL_KUL&collectionId=81411248550001488} & \url{https://doi.org/10.5281/zenodo.7762478}\\
    \hline
    Anjou Bible & \url{https://kuleuven.limo.libis.be/discovery/fulldisplay?docid=alma9983846510101488&context=L&vid=32KUL_KUL:KULeuven&lang=en} & \url{https://doi.org/10.5281/zenodo.7762363}\\
    \hline
    Postcards Belgium & \url{https://kuleuven.limo.libis.be/discovery/collectionDiscovery?vid=32KUL_KUL:KULeuven&inst=32KUL_KUL&collectionId=81411181930001488} & \url{https://doi.org/10.5281/zenodo.7769559}\\
    \hline
    Wartime Posters & \url{https://kuleuven.limo.libis.be/discovery/collectionDiscovery?vid=32KUL_KUL:KULeuven&inst=32KUL_KUL&collectionId=81411182030001488} & \url{https://doi.org/10.5281/zenodo.7761872}\\
     \hline
    Historical Censuses Belgium & \url{https://kuleuven.limo.libis.be/discovery/collectionDiscovery?vid=32KUL_KUL:KULeuven&inst=32KUL_KUL&collectionId=81423334580001488} & \url{https://doi.org/10.5281/zenodo.7764048}\\
     \hline
    Academic Collection of the Old University of Leuven & \url{https://kuleuven.limo.libis.be/discovery/collectionDiscovery?vid=32KUL_KUL:KULeuven&inst=32KUL_KUL&collectionId=81411248640001488} & \url{https://doi.org/10.5281/zenodo.7762634}\\
  \bottomrule
\end{tabular}
}
\end{table}

Parameters for creating datasets depend on the context, the collection content, target users and intended use~\cite{padilla_thomas_2019_3152935}. Six datasets (see Table \ref{tab:leuvendatasets}) were created as part of the preparation for the a hackathon aimed at researchers and postgraduate students from within KU Leuven \cite{kuleuvenlibraries}. The hackathon organisation was a collaborative effort of KU Leuven Libraries and the university's Faculty of Arts\footnote{https://www.arts.kuleuven.be/digitalhumanities/english}. Considering this context and as this was the library's first endeavor in creating collections as data, it was decided to (at least temporarily) offer access to the datasets (item 4, \emph{use a public platform to publish the dataset}) through KU Leuven's internal data portal. \emph{Share examples of use} (item 5) and \emph{include your dataset in collaborative edition platforms} (item 8) were irrelevant to this context as participants were expected to work directly and independently with the source data. Regarding \emph{-providing a clear license allowing reuse of the dataset without restrictions} (item 1), it was decided to only include resources in the public domain in order to allow for full reuse by participants outside of the hackathon. The Public Domain mark was provided on a resource-level in the descriptive metadata of the resources included in the dataset. \emph{Add a terms of use} (item 11) was satisfied by including a statement in the hackathon Code of Conduct.

To start, the metadata and data were identified and extracted from their respective repositories. The dataset was subsequently structured according to \emph{give structure to the dataset} (item 6): each dataset contained a separate folder for each resource within which there were subfolders for each of the representations of these resources, e.g. a folder for page-level OCR data, for page-level jp2 images, and for PDF. On the level of the resource, a json manifest was included describing the resource. At the dataset level, the descriptive metadata was included as a xml metadata dump and as a partially cleaned csv and excel. A final csv was also provided, revealing the concordance between all the files in the dataset. 

The full dataset was uploaded to the internal KU Leuven active data portal ManGO (item 4, \emph{use a public platform to publish the dataset}), where hackathon participants could access the data, execute downloads or (providing the necessary infrastructure to work with the large datasets) connect the high performance computing infrastructure to the data through an API (item 9, \emph{offer an API to access the repository}). In ManGO dataset metadata was added but for query purposes only. Documentation to each of the datasets included full information on the dataset structure, the descriptive metadata model, and some information on the level of the physical collection on the basis of which these datasets were created. 

The library plans to further develop the datasets, improving there (re-)useability by including a Terms of use on a dataset level (item 11, \emph{add a terms of use}), investigating possible locations to store and access the datasets for non-KU Leuven users (items 4, 9 and/or 10) as well as citation information which is so far lacking as the data is non-accessible to non-KU Leuven users (item 2, \emph{provide a suggestion of how to cite your dataset}), and improving the structure and completeness of the datasets' documentation (item 3, \emph{include documentation about the dataset}). It will also include the DOI of the hackathon team posters to inspire potential users (item 5, \emph{share examples of use as additional documentation}), and investigate and develop a metadatamodel for dataset metadata (item 7, \emph{provide machine-readable metadata}). The library has already pushed other data to a collaborative editing platform such as Wikidata and considers doing the same for these datasets.\footnote{See, for example, \url{https://www.wikidata.org/wiki/Q112958007} and \url{https://commons.wikimedia.org/wiki/Category:Glass_diapositives_Egyptology,_KU_Leuven_Libraries}} Yet, due to a lack of personnel, this will be postponed to a later date. 

As a whole, the datasets directed the preparation phase by providing and concise overview of which actions lead to collections as data, allowing for timely reflection on priorities. It now also supports decisions regarding next steps to take. 

\subsection{Towards a Collections as data platform at KBR, the Royal Library of Belgium}
\label{sec:kbr}

Inspired by the Collections as data movement \cite{padilla_thomas_2019_3066237,padilla_thomas_2019_3152935}, in 2020, KBR, the Royal Library of Belgium, embarked on a 48 month project\footnote{DATA-KBR-BE is financed by the Belgian Science Policy Office (Belspo) as part of the Belgian Research Action through Interdisciplinary Networks, BRAIN 2.0 programme. Originally foreseen as a 24-month project, in February 2022, the project was extended until 15th March 2024.} called DATA-KBR-BE \footnote{\url{https://www.kbr.be/en/projects/data-kbr-be/}} (2020-2024). The aim of the project is to optimise KBR’s ICT infrastructure to stimulate sustainable data-level access to KBR’s digitised and born-digital collections for digital humanities research. A key output of the project is to design and implement an Open Data Platform (data.kbr.be), for publishing KBR’s collections as data datasets. 

Work on conceptualising the future DATA-KBR-BE platform began at an early stage in the project, during an initial \textit{Brainstorming Workshop} held in November 2020. It was clear from the outset that a researcher-centred and iterative approach was needed to gather requirements for the design and development of the DATA-KBR-BE platform. A first important step in this process was to review some of the existing library data platforms, such as the national libraries of Luxembourg ,\footnote{\url{https://data.bnl.lu/data/historical-newspapers/}} the Netherlands,\footnote{\url{https://lab.kb.nl/products/product_type/dataset}} Scotland\footnote{\url{https://data.nls.uk}} and The British Library.\footnote{\url{https://data.bl.uk/}} Questions such as: \textit {What data is offered? How? What format? What did people like, dislike about the platforms which were explored?} The outcomes of this workshop were used as the basis for iteratively developing a checklist of needs. The emergence of the “Checklist to publish Collections as data in GLAM Institutions” introduced in Section \ref{sec:checklist} provided the project team with an ideal framework to help structure the development of the functional and technical requirements for the platform. 

To prepare for the webinar in October 2022 \cite{candela_gustavo_2022_7789480},\footnote{\url{https://glamlabs.io/events/collections-data/}} an initial analysis of the checklist was undertaken to assess which checklist items were most relevant for the DATA-KBR-BE project. Initially, \textit{develop a portal page} (item 10) and \textit {give structure to the dataset} (item 6) were identified as the most relevant items for the project team, as our aim was to develop the DATA-KBR-BE platform and that we would need to understand how to structure the datasets that will be published there. However, it soon became relevant that many, if not all, the checklist items would support the development of the DATA-KBR-BE platform. For example, \textit{provide a suggestion of how to cite your dataset} (item 2) and \textit {provide a license allowing reuse of the dataset} (item 1) were quickly seen as essential. 

\begin{figure*}
\includegraphics[scale=0.3]{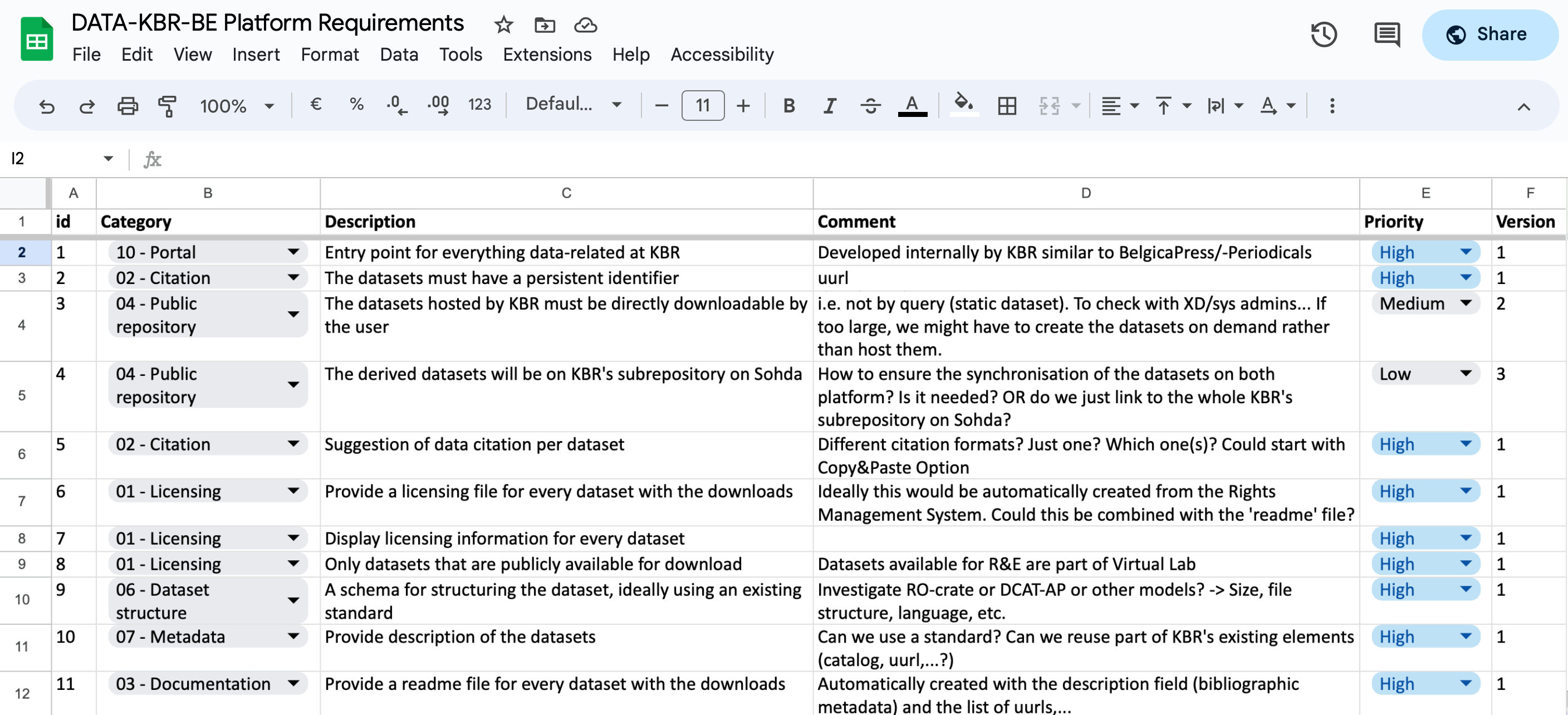}
\caption{A collaborative spreadsheet for capturing the technical and functional requirements for the DATA-KBR-BE platform}
\label{fig:kbr}
\end{figure*}

To use the checklist more systematically, a collaborative spreadsheet was designed to capture each of the functional and technical requirements for the DATA-KBR-BE platform, as shown in Figure \ref{fig:kbr}. Column B, is used for categorising each of the requirements based on the checklist list, e.g. requirement 1: \textit {entry point for everything data-related at KBR}, has been categorised in relation to checklist item \textit {develop a portal page} (item 10). This approach enabled us to: a) group the requirements by category, b) to ensure that our requirements analysis was as exhaustive as possible by considering each of the checklist items, and c) to provide feedback to the International GLAM Labs Community to further improve the checklist. 

When reviewing the checklist items, the difference between \textit {use a public platform to publish the dataset} (item 4) and \textit {develop a portal page} (item 10) was not initially clear without further explanation. We were also not sure about \textit {include your dataset in a collaborative edition platform} (item 8) and also how it relates to items 4 and 10. Additionally, \textit {Provide machine-readable metadata} (item 7) caused quite some discussion in the DATA-KBR-BE project team. For example, what about human-readable metadata? What does machine-readable mean in this context? Why is it prioritised? Is this descriptive metadata? Are any particular standards recommended? How does this relate to structural metadata? Is this covered under \textit {give structure to the dataset?} (item 6). The training and documentation aspects of the checklist \textit {include documentation about the dataset} (item 3) and \textit {share examples of use as additional documentation} (item 5) were both seen as very relevant to the development of the DATA-KBR-BE platform. However, they would likely be added following the initial development of the platform itself. Finally, \textit {offer an API access to your repository} (item 9) was out of the scope of the current DATA-KBR-BE project, and will be addressed in a follow-up project, the KBR Virtual Lab. 

In conclusion, the “Collections as Data Checklist” was a valuable tool as it helped ensure that the DATA-KBR-BE project team had considered as many of the aspects of the checklist as possible when developing our Collections as data platform.

\subsection{Discussion}

While various institutions have made available their collections using APIs, there are still barriers hindering the adoption of the Collections as data principle, such as the lack of resources as well as the desire to make the collections easily available to a broad user group by means of simple access and downloads. The GLAM datasets which were selected for assessment in this article present some similarities but also some differences, e.g., the type of content, the formats and standards used for digital delivery, how they can be accessed, the licensing, and the documentation provided.

The checklist is informed by the issues and needs identified within the literature review and is complemented by the contributions of the practitioners who considered all the items included in the checklist relevant. In general, the practitioners observed a balance between simplicity and depth of practice. Some of them remarked that each of the items requires a different degree of maturity and prioritisation, which in some cases necessitates joint efforts by the community.

With regard to the application of the checklist as an assessment tool, and taking into account the wide variety of datasets provided by the GLAM institutions, the results obtained after the application of the checklist may differ amongst adopters of this approach. Initial results showed that the checklist is useful for identifying which aspects are relevant for a particular institution and, to some extent, easy to apply when making available datasets for computational use. In general, we observed that there is no order when applying the items in the checklist. Rather, as the case of KU Leuven Libraries demonstrates, priorities depend on the context, the content, the intended use and target users of the dataset. Furthermore, the checklist further can facilitate the development of infrastructures related to collections as data, as shown in the case of the DATA-KBR-BE platform. 

In general, future work based on the items in the checklist is a common goal across the institutions wishing to make their collections available as data. Regarding the sharing of examples of practical implementations, the use of Jupyter Notebooks is increasing across organisations that are working in this area. For example, the British Library plans to improve reuse examples and their related tools on data.bl.uk in due course. Concerning the structure of the datasets, institutions are interested in improvements which can enhance the user experience. 

While the institutional journeys into the delivery of collections as data differ and are not yet taking place in all institutions in the GLAM sector, an additional layer of complexity in the computational use of cultural data which needs to be accommodated is the development of the common European data space for cultural heritage, the European Cultural Heritage Cloud and other data spaces which will be using GLAM data, e.g. the European Open Science Cloud and large research infrastructures in the digital humanities. There is an ongoing effort to identify use cases for the data space for cultural heritage, and it would be helpful to coordinate this work with the future refinement of the checklist.

\section{Conclusions}
Over the past few years, there has been a growing interest in making available the digital collections published by GLAM organisations for computational use. 

Based on previous work, we defined a methodology described in Section~\ref{sec:framework} to build a checklist for the publication of collections as data. Our evaluation showed several examples of application that can be useful to encourage other institutions to publish their digital collections for computational use.

Future work to be explored includes the improvement of the methodology by including additional features such as carbon footprint assessment, ethical issues and quality, as well as the inclusion of additional collections as data provided by organisations as use cases.

\begin{acks}
The authors would like to thank the International GLAM Labs Community and all the GLAM institutions that contributed to the survey. We would like to thank Ariadna Matas from Europeana Foundation for her suggestions regarding licenses and tools in Section \ref{subsec:license}.
\end{acks}

\bibliographystyle{ACM-Reference-Format}
\bibliography{cad-labs}

\appendix

\section{List of studies}
\label{app:review}

Tables \ref{tab:review} and \ref{tab:review2} presents the list of primary studies analysed as a literature review on the use of Collections as data in GLAM institutions.

\begin{table}
  \caption{List of primary studies analysed as a literature review.}
  \label{tab:review}
  \scalebox{0.59}{
  \begin{tabular}{p{9cm}p{3.5cm}cp{3.5cm}}
    \toprule
    Title&Origin&Type&Description\\
    \midrule
    50 Things --- Always Already Computational: Collections as data & Collections as data project & Report & Best practices\\
    \hline
    A benchmark of Spanish language datasets for computationally driven research & Journal of Information Science & Research article & Best practices\\
    \hline
    A checklist for the evaluation of software process line approaches & Journal Information \& Software Technology & Research article & Checklist\\
    \hline
    A checklist recipe: making species data open and FAIR & Database J. Biol. Databases Curation & Research article & Checklist\\
    \hline
    A sustainable future for our digital assets 2022 – 2027 & Digital Preservation Coalition & Report & Strategy\\
    \hline
    Artificial Intelligence Roadmap 2021-2026 & Bibliothèque nationale de France & Report & Strategy\\
    \hline
    Assessing the Impact of OCR Quality on Downstream NLP Tasks & ICAART 2020 & Research article & Quality\\
    \hline
    BnL’s Technical Requirements & Bibliothèque nationale du Luxembourg & Report & Requirements for digitization projects\\
    \hline
    British Library Datasets & British Library & Datasets & Data publication\\
    \hline
    Checklist for a Data Management Plan. v.4.0. & Digital Curation Centre & Report & Checklist\\
    \hline
    Checklist for Validating Trustworthy AI & Conference & Research article & Checklist\\
    \hline
    Creating Library Linked Data with Wikibase: Lessons Learned from Project Passage & OCLC Research & Report & Best practices\\
    \hline
    Data & Biblioteca Nacional de España & Datasets & Data publication\\
    \hline
    Data & Data Foundry & Datasets & Data publication\\
    \hline
    Data outputs & Australian Cultural Data Engine & Report & Data publication\\
    \hline
    Datasheets for datasets & Communications of the ACM & Research article & Data publication\\
    \hline
    Digital Libraries, Intelligent Data Analytics, and Augmented Description: A demonstration project & Library of Congress & Report & Best practices\\
    \hline
    Digital Scholarship at the Library of Congress & Library of Congress & Report & Best practices\\
    \hline
    Discovering emerging topics in textual corpora of galleries, libraries, archives, and museums institutions & Jasist Journal & Research article & Examples of use\\
    \hline
    Environmental scan: Artificial Intelligence, cultural heritage and the National Library of Scotland & NLS & Report & Strategy\\
    \hline
    Exploring Data Provenance in Handwritten Text Recognition Infrastructure: Sharing and Reusing Ground Truth Data, Referencing Models, and Acknowledging Contributions. Starting the Conversation on How We Could Get It Done & Zenodo & Research article & Best practices\\
    \hline
    Evaluating the quality of linked open data in digital libraries & Journal of Information Science & Research article & Quality\\
    \hline
    Facilitating data-level access to KBR’s digitised and born-digital collections for digital humanities research & Royal Library of Belgium & Website & Project description\\
    \hline
    Final Report --- Always Already Computational: Collections as data & Collections as data project & Report & Outcomes of the project\\
    \hline
    Foundations for the Future. The British Library’s Collection Metadata Strategy 2019-2023 & British Library & Report & Strategy\\
    \hline
    From collection search to Collections as data & Tim Sherratt & Report & Best practices\\
    \hline
    Harvard Art Museums API & Harvard Art Museums & Website & Technical documentation\\
    \hline
    Historical Newspapers & Bibliothèque nationale du Luxembourg & Datasets & Data publication\\
    \hline
    Humans-in-the-Loop Recommendations report & Library of Congress & Report & Best practices\\
    \hline
   
    LIBER Europe Strategy 2018-2022 & LIBER & Report & Strategy\\
    \hline
    Machine learning and libraries: a report on the state of the field & Library of Congress & Report & Best practices\\
    \hline
    Machine Learning + Libraries Summit Event Summary Library of Congress & Report & Best practices\\
    \hline
    Migration of a library catalogue into RDA linked open data & Semantic Web Journal & Research article & Data publication \\
    \hline
    Next generation of metadata & OCLC Research & Report & Best practices\\
    \hline
    On art authentication and the Rijksmuseum challenge: A residual
               neural network approach & Rijksmuseum & Research article & Experiment\\
    \hline
    Open Data Plan & National Library of Scotland & Open Data Plan & Best practices\\
    \hline
    Open a GLAM Lab & International GLAM Lab community & Book & Best practices\\
    \hline
    Responsible Operations: Data Science, Machine Learning, and AI in Libraries & OCLC Research & Report & Best practices\\
    \hline
    Reusing digital collections from GLAM institutions & Journal of Information Science & Research article & Examples of use\\
    \hline
    RLUK Strategy 2022-2025 & Research Libraries UK & Report & Strategy\\
    \hline
    Strategic Plan 2020-2023 & National and State Libraries Australasia & Report & Strategy\\
    \hline
    Strategy 2020-2025 & Europeana & Report & Strategy\\
  \bottomrule
\end{tabular}
}
\end{table}


\begin{table}
  \caption{List of primary studies analysed as a literature review (continued from previous page)}
  \label{tab:review2}
  \scalebox{0.59}{
  \begin{tabular}{p{9cm}p{3.5cm}cp{3.5cm}}
    \toprule
    Title&Origin&Type&Description\\
    \midrule
    Supporting Ethical Data Research: An Exploratory Study of Emerging Issues in Big Data and Technical Research & Data \& Society & Report & Best practices\\
    \hline
    The "Collections as ML Data" Checklist for Machine Learning and Cultural Heritage & ArXiv & Research article & Checklist\\
    \hline
    The Museum of Modern Art (MoMA) API is a REST service & MoMA & Website & Technical documentation\\
     \hline
    The Newspaper Navigator Dataset & ArXiv & Research article & Data publication\\
    \hline
    The Rijksmuseum collection as Linked Data & Semantic Web Journal & Research article & Data publication\\
    \hline
    Validating 126 million MARC records & DATeCH 2019 & Research article & Quality\\
    \hline
    WarSampo knowledge graph: Finland in the Second World War as Linked Open Data & Semantic Web Journal & Research article & Data publication\\
  \bottomrule
\end{tabular}
}
\end{table}

\end{document}